\documentclass[usenatbib,usegraphicx]{mn2e}

\def\references{\bibliographystyle{mn2e}\bibliography{draft}}

\usepackage{amsbsy} 
\usepackage{latexsym}
\def\jnl@style{\rm}
\def\reffmtPrivate@jnl#1{{\jnl@style#1}}
\def\aj{\reffmtPrivate@jnl{AJ}}                   
\def\araa{\reffmtPrivate@jnl{ARA\&A}}             
\def\apj{\reffmtPrivate@jnl{ApJ}}                 
\def\apjl{\reffmtPrivate@jnl{ApJ}}                
\def\apjs{\reffmtPrivate@jnl{ApJS}}               
\def\ao{\reffmtPrivate@jnl{Appl.~Opt.}}           
\def\apss{\reffmtPrivate@jnl{Ap\&SS}}             
\def\aap{\reffmtPrivate@jnl{A\&A}}                
\def\aapr{\reffmtPrivate@jnl{A\&A~Rev.}}          
\def\aaps{\reffmtPrivate@jnl{A\&AS}}              
\def\azh{\reffmtPrivate@jnl{AZh}}                 
\def\baas{\reffmtPrivate@jnl{BAAS}}               
\def\jrasc{\reffmtPrivate@jnl{JRASC}}             
\def\memras{\reffmtPrivate@jnl{MmRAS}}            
\def\mnras{\reffmtPrivate@jnl{MNRAS}}             
\def\pra{\reffmtPrivate@jnl{Phys.~Rev.~A}}        
\def\prb{\reffmtPrivate@jnl{Phys.~Rev.~B}}        
\def\prc{\reffmtPrivate@jnl{Phys.~Rev.~C}}        
\def\prd{\reffmtPrivate@jnl{Phys.~Rev.~D}}        
\def\pre{\reffmtPrivate@jnl{Phys.~Rev.~E}}        
\def\prl{\reffmtPrivate@jnl{Phys.~Rev.~Lett.}}    
\def\pasp{\reffmtPrivate@jnl{PASP}}               
\def\pasj{\reffmtPrivate@jnl{PASJ}}               
\def\qjras{\reffmtPrivate@jnl{QJRAS}}             
\def\skytel{\reffmtPrivate@jnl{S\&T}}             
\def\solphys{\reffmtPrivate@jnl{Sol.~Phys.}}      
\def\sovast{\reffmtPrivate@jnl{Soviet~Ast.}}      
\def\ssr{\reffmtPrivate@jnl{Space~Sci.~Rev.}}     
\def\zap{\reffmtPrivate@jnl{ZAp}}                 
\def\nat{\reffmtPrivate@jnl{Nature}}              
\def\iaucirc{\reffmtPrivate@jnl{IAU~Circ.}}
\def\aplett{\reffmtPrivate@jnl{Astrophys.~Lett.}}
\def\apspr{\reffmtPrivate@jnl{Astrophys.~Space~Phys.~Res.}}
\def\bain{\reffmtPrivate@jnl{Bull.~Astron.~Inst.~Netherlands}}
\def\fcp{\reffmtPrivate@jnl{Fund.~Cosmic~Phys.}}
\def\gca{\reffmtPrivate@jnl{Geochim.~Cosmochim.~Acta}}
\def\grl{\reffmtPrivate@jnl{Geophys.~Res.~Lett.}}
\def\jcp{\reffmtPrivate@jnl{J.~Chem.~Phys.}}      
\def\jgr{\reffmtPrivate@jnl{J.~Geophys.~Res.}}    
\def\jqsrt{\reffmtPrivate@jnl{J.~Quant.~Spec.~Radiat.~Transf.}}
\def\memsai{\reffmtPrivate@jnl{Mem.~Soc.~Astron.~Italiana}}
\def\nphysa{\reffmtPrivate@jnl{Nucl.~Phys.~A}}
\def\physrep{\reffmtPrivate@jnl{Phys.~Rep.}}
\def\physscr{\reffmtPrivate@jnl{Phys.~Scr}}
\def\planss{\reffmtPrivate@jnl{Planet.~Space~Sci.}}
\def\procspie{\reffmtPrivate@jnl{Proc.~SPIE}}
\def\adass{\reffmtPrivate@jnl{Astron.~Data~Software~and~Systems}}
\def\amasm{\reffmtPrivate@jnl{Am.~Astron.~Soc.~Meeting}}

\newcommand{\be}[1]{\begin{equation}\label{#1}}
\newcommand{\ee}{\end{equation}}
\newcommand{\msun}{{\;M_{\odot}}}
\newcommand{\rmd}{\mathrm{d}}

\newcommand{\bsy}{\boldsymbol}

\newcommand{\mrm}{\mathrm}

\newcommand{\bea}[1]{\begin{eqnarray}\label{#1}}
\newcommand{\eea}{\end{eqnarray}}

\newlength{\onefig}
\newlength{\twofig}
\newlength{\figshift}
\newcommand{\mytwofigure}[6]{
\begin{figure*}
\includegraphics[width=\onefig,angle=0]{#1}
\includegraphics[width=\onefig,angle=0]{#2}
\caption[]{#5}
\label{#6}
\end{figure*}
}

\newcommand{\UAB}{{$U_{\mathrm{AB}}$}}
\newcommand{\RAB}{{$R_{\mathrm{AB}}$}}
\newcommand{\URAB}{{$(U-R)_{\mathrm{AB}}$}}
\newcommand{\UABlim}{{$U_{\mathrm{AB,lim}}$}}
\newcommand{\RABlim}{{$R_{\mathrm{AB,lim}}$}}

\title[Depletion due to the cluster lens CL0024+1654]{Depletion of
background galaxies due to the cluster lens CL0024+1654: $U$ and $R$ band observations}

\author[\"{O}.~E.~R\"{o}gnvaldsson et al.]{\"{O}.~E.~R\"{o}gnvaldsson,$^1$\thanks{email: ossi@nordita.dk}
T.~R.~Greve,$^2$ J.~Hjorth,$^2$
\newauthor
E.~H.~Gudmundsson,$^3$ V.~S.~Sigmundsson,$^3$ P.~Jakobsson,$^{2,3}$
\newauthor
A.~O.~Jaunsen,$^4$
L.~L.~Christensen,$^5$
E.~van Kampen$^6$ and A.~N.~Taylor$^6$ \\
$^1$NORDITA, Blegdamsvej 17, DK-2100 Copenhagen {\O}, Denmark \\
$^2$Astronomical Observatory, Univeristy of Copenhagen,
Juliane Maries Vej 30, DK-2100 Copenhagen {\O}, Denmark \\
$^3$Science Institute, University of Iceland,
       Dunhaga 3, IS-107 Reykjavik, Iceland \\
$^4$Institute of Theoretical Astrophysics,
          Postboks 1029 Blindern, N-0135 Oslo, Norway \\
$^5$ST-ECF/ESO,
           Garching, Germany \\
$^6$Institute for Astronomy,
           Blackford Hill, Edinburgh EH9 3HJ, UK}
\date{Accepted \ldots
      Received 2000 July 17; revised 2000 August 30}

\newcommand{\Mythesaurus}{}
\newcommand{\Mymaketitle}{\maketitle}
\newcommand{\Myacknowledgements}{\section*{Acknowledgements}}
\newcommand{\Myacknowledgementsend}{}

\begin{document}

\Mythesaurus{}
\Mymaketitle

\def\rE{r_{\mrm{E}}}
\def\rEp{\rE'}
\def\zM{\langle z_{\mrm{S}} \rangle}

\begin{abstract}
We have obtained $U$ and $R$ band observations of the depletion of background 
galaxies due to the gravitational lensing of the galaxy cluster 
CL0024+1654 ($z=0.39$).
The radial depletion curves show a significant depletion in both bands
within a radius of
$40\arcsec-70\arcsec$~from the cluster center. This is the first time
depletion is detected in the $U$ band. This gives independent evidence for
a break in the slope of the $U$ band luminosity function at faint
magnitudes.
The radially averaged $R$ band depletion curve is broader and deeper
than in the $U$ band.
The differences can be attributed to the wavelength
dependence of the slope of the luminosity function and to the
different redshift distribution of the objects probed in the two
bands. We estimate the Einstein
radius, $\rE$, of a singular isothermal sphere lens model using maximum
likelihood analysis.
Adopting a slope of the number counts of $\alpha=0.2$ and using
the background density found beyond $r=150\arcsec$ we
find $\rE = 17\arcsec \pm
3\arcsec$ and $\rE = 25\arcsec \pm 3\arcsec$ in the $U$ and $R$ band,
respectively. When combined with the redshift of the single
background galaxy at $z=1.675$ seen as four giant arcs around
$30\arcsec$ from the cluster center, these values indicate a median
redshift in the range $\zM \approx 0.7$ to $1.1$ for the
\UAB$\geq 24$ mag and \RAB$\geq 24$ mag populations.

\end{abstract}


\section{Introduction}

Gravitational lensing by clusters of galaxies affects the
apparent distribution of background galaxies on the sky, provided that
the slope of the galaxy number counts differs from a critical value
corresponding to a balance between the dilution due to local stretching
of the background sky and the density enhancement due to magnification
of faint sources above the observed magnitude limit. Normally, the
slope levels off at faint magnitudes, causing an apparent depletion
of galaxies behind strong gravitational lenses relative to the field.
This depletion signal is, under fairly general conditions, directly
related to the
magnification of the lens, which in turn may be used to find its
surface mass density. \cite{Broadhurst+Taylor+Peacock;1995} first pointed
out the potential of this aspect of the magnification bias to estimate
cluster masses and density profiles, using the so-called number count
method.
This method was first used for absolute mass estimation by
\cite{Taylor+Dye+Broadhurst+;1998}, who also discuss various
observational issues. Models for relating the magnification to the
surface mass density (or convergence), going beyond the weak lensing
approximation, were considered by \cite{vanKampen;1998} and
\cite{Dye+Taylor;1998}. Such models, which take into account the
presence of a shear term in the magnification, are of great importance in
combination with the number count method, since the shear pattern
can not be measured from the background densities alone.
\cite{Schneider+King+Erben;2000} have
compared the number count method to shear measurements,
using maximum likelihood analysis. They find that the number count
method does a better job of estimating cluster density profiles,
provided that the unlensed number density of galaxies is known to a
good precision. 
\cite{Bezecourt+Kneib+Soucail+;1999} have investigated the
wavelength dependence of depletion curves, while
\cite{Mayen+Soucail;2000} investigate how depletion curves
depend on the cluster density profile. They also investigate how the
filter bands used in observations affect the results
through different sampling of the background population.
\cite{Gray+Ellis+Refregier+;2000} show that the effects of
incompleteness on model parameters estimated with maximum likelihood
methods (\cite{Schneider+King+Erben;2000}) may be neglected, at least under the
standard assumption that the intrinsic luminosity function is a power
law.

A few other methods based on
gravitational lensing have been applied to cluster mass
estimation. \cite{Dye+Taylor+Thommes+;2000} have studied the effects
of the lens magnification on the luminosity function of background
population, and applied this method to Abell 1689. Weak
shear analysis, where the average distortion of the shape of
background galaxies is used to estimate the shear field of the lens,
has been applied to a number of clusters
(e.g.~\cite{Smail+Dickinson;1995,Luppino+Kaiser;1997,Cowe+Luppino+;1998,Hoekstra+Franx+Kuijken;2000,Joffre+Fischer+;2000,Athreya+Mellier+Waerbeke+;2000},
and references therein).
Strong lensing effects
have also been used to probe the density distribution in clusters
(e.g.~\cite{Tyson+Kochanski+DellAntonio;1998,Broadhurst+Huang+Frye+;2000}),
and statistical studies of arclets are being pursued
(e.g.~\cite{Bezecourt+Kneib+Soucail+;1999}). The theoretical and
observational aspects of
gravitational lensing and its applications have recently been
reviewed in \cite{Mellier;1999} and \cite{Bartelmann+Schneider;2000}.

To date, depletion of background galaxies has been observed in
connection with a
handful of clusters. \cite{Fort+Mellier+;1997} were the first to
report on depletion in the $B$ and $I$ band behind CL0024+1654. Their
observations were however plagued by a low density of objects and an
uncertain background estimate as discussed by \cite{vanKampen;1998}.
\cite{Taylor+Dye+Broadhurst+;1998} found depletion in the $I$ band
in Abell 1689, after careful removal of cluster and foreground objects
using additional color information.
\cite{Mayen+Soucail;2000} find depletion of faint $B$, $V$, $R$, and
$I$ objects towards MS1008--1224. This cluster can possibly be used
to investigate how
background clustering affects the depletion analysis, since
\cite{Athreya+Mellier+Waerbeke+;2000} have tentatively detected another cluster
in the background (at $z=0.9$). Finally,
\cite{Gray+Ellis+Refregier+;2000} have explored depletion at
near-infrared wavelengths in Abell 2219. Depletion analysis in the
infrared (IR) benefit from a low slope of the number counts already at
moderately faint IR magnitudes, but the low background number density
may outweigh the benefits.

In our work, we apply the number count method in the $U$ and $R$ band
to the rich cluster
CL0024+1654 (z=0.39), first observed to be a strong gravitational lens
by \cite{Koo;1988}.
The $U$ filter is well suited for depletion studies,
provided that the observations are deep enough to probe galaxies
fainter than \UAB$\approx 25.75$ where the slope of the luminosity
function in the $U$ band levels off
(\cite{Williams+Blacker+Dickinson+;1996,Pozzetti+Madau+;1998},
but see e.g.~\cite{Hogg+Pahre+;1997,Volonteri+Saracco+;2000} for
evidence against this),
while the slope in the $R$ band
is low enough to accommodate depletion studies given sufficient
number of background objects. With sufficiently deep observations, depletion
analysis in these two widely separated bands will also reflect the
different sampling of the background redshift distribution.
The large Einstein radius of CL0024+1654, inferred from the position
of giant arcs around
$30\arcsec$ from the cluster center, and previous detection of depletion
in the $I$
and $B$ band (\cite{Fort+Mellier+;1997}) make this cluster a good
candidate for
depletion studies. Moreover, its mass and mass distribution have been
studied carefully with other methods.

From the redshift catalog compiled by
\cite{Dressler+Gunn+Schneider;1985}, \cite{Schneider+Dressler+Gunn;1986}, and
\cite{Dressler+Gunn;1992}, a velocity dispersion of
$\sigma_v = 1200$ km s$^{-1}$ is found, resulting in a mass of roughly $7
\times 10^{14} h_{50}^{-1} \msun$ within a radius of $0.5 h_{50}^{-1}$
Mpc. (The Hubble parameter is $H_0 = 50 h_{50}$ km/s/Mpc.)
A much larger sample of redshifts in the cluster field has
recently been obtained by
\cite{Dressler+;1999} and \cite{Czoske+Soucail+Kneib+;1999}. Analysis
of this data leads to a
considerably lower velocity dispersion, $\sigma \approx 700$ km s$^{-1}$, for
the 227 cluster objects identified. The estimated mass is consequently
lower. For example \cite{Czoske+Soucail+Kneib+;1999} find a mass of $1.4 \times
10^{14} h_{50}^{-1} \msun$ within $0.5 h_{50}^{-1}$ Mpc.

From a detailed inversion of the lens using mainly the giant arcs,
which are the multiple images of a single background galaxy,
\cite{Tyson+Kochanski+DellAntonio;1998} find the central mass
distribution to be very relaxed, with no signs of substructure within
$220 h_{50}^{-1}$ kpc. On the other hand,
\cite{Broadhurst+Huang+Frye+;2000} find
evidence for substructure on the cluster core scale. Their mass
estimate for the innermost $220 h_{50}^{-1}$ kpc is
$2.6 \times 10^{14} h_{50}^{-1} \msun$, or somewhat lower than the
$3.3 \times 10^{14} h_{50}^{-1} \msun$ estimated by
\cite{Tyson+Kochanski+DellAntonio;1998}. Their model does however
predict an unrealistically high velocity dispersion for the cluster
(\cite{Shapiro+Iliev;2000}).

On a much larger scale, 
\cite{Bonnet+Mellier+Fort;1994} using weak shear analysis find
a mass of $4 \times 10^{15} h_{50}^{-1} \msun$ within a
radius of $3 h_{50}^{-1}$ Mpc.
Analysis of X-ray observations by 
\cite{Bohringer+Soucail+Mellier+;2000} and \cite{Soucail+Ota+Bohringer+;2000}
shows a discrepancy of about a factor $1-3$ between X-ray and
lensing mass, similar to what is found in many other clusters. The
difference is most probably attributable to simplified assumptions
 about the dynamical state of the cluster. Finally, \cite{vanKampen;1998} used
the depletion observations by \cite{Fort+Mellier+;1997} to
constrain the mass of the cluster, noting that a better estimate of
the unlensed background density was needed to give reliable results.

In order to constrain the mass distribution of the cluster using our data,
additional information is needed on various properties of
the background sources.
The unlensed average density, redshift distribution, and the
slope of the number counts of the background sources all play
a role in relating the observed depletion to the convergence of the lens.
In this paper, we present the first $U$ band observations of depletion
and apply simple parametric modeling to our $U$ and $R$ band data to
investigate the dependence of the model on the a priori parameters.
A more detailed investigation of the mass distribution
implied by our data is being undertaken (van Kampen et al., in preparation).

The rest of this paper is organized as follows. Basic description of the
data is given in Section~\ref{sec:data}. The object extraction and
analysis of the resulting catalogs is discussed in Section
\ref{sec:analysis}, and the results are discussed and summarized in Section
\ref{sec:discuss+conclude}.

\section{The data}
\label{sec:data}
\subsection{Data acquisition and reduction}
The data were obtained with the ALFOSC instrument on the Nordic
Optical Telescope in August 1999. The field-of-view was
6.5$\Box\arcmin$ with a pixel size of $0.188\arcsec$.
The total integration time in the
(Cousins) $U$ band was 37 ksec and 8.7 ksec in the (Cousins) $R$ band with
average seeing of $1.1\arcsec$ and $1.0\arcsec$, respectively. The formal $3\sigma$
detection limits in the final drizzled images are \UABlim =
27.2 mag and \RABlim = 27.5 mag. (We use
AB magnitudes throughout, using a simple shift of +0.71 mag and +0.199
mag to convert from the observed Cousins $U$ and $R$ magnitudes,
respectively (\cite{Fukugita+Shimasaku+Ichikawa;1995}).

Standard bias and flat field corrections were made to the data before
stacking the individual frames using tools from the
{\tt ditherII} package
(\cite{Fruchter+Hook;1997,Fruchter+Mutchler;1998,Hook+Pirzkal+Fruchter;1999})
in {\tt IRAF}.
Shifts were estimated by
cross-correlating images and cosmic rays masked out using the blotting
technique as described by \cite{Fruchter+Mutchler;1998}. The original
pixel scale was retained and no attempt
was made to correct for field distortions.

\subsection{Photometry and completeness}
The calibration against photometric standards (mainly in M92
(\cite{Davis:M92}) and a few standards from
\cite{Landolt;1983}) gives formal errors of $\sigma_{\Delta R} = 0.02$
mag and $\sigma_{\Delta U} = 0.05$ mag. Typical photometric errors
estimated by the detection software are between $0.1-0.3$ mag for the
magnitude ranges $24 \leq (R,U)_{\mrm{AB}} \leq 27$. (This is the
magnitude range used
below to select background objects for the depletion analysis.)

In order to estimate the level of completeness as a function of
magnitude in our data, 300 artificial objects were added to the science
frames using the {\tt artdata}
task in {\tt IRAF}. Completeness was estimated as the ratio between detected
synthetic objects and the number of synthetic objects added. In
order to improve the statistics this procedure was repeated 100 times,
each time adding objects at random positions. The final average
completion curves indicate that our sample is 50\% complete at
\UAB = $26.7$ mag and \RAB = $26.5$.

The fraction of detections due to spurious noise peaks will steadily
increase towards faint flux levels. To investigate this,
frames containing only noise were generated
and using SExtractor (\cite{Bertin+Arnouts;1996}), 700
apertures were dropped onto the frames. To ensure that the noise-only frames
would mimic the noise characteristics of the science frames as
closely as possible, they were convolved with the noise correlation
function found in the science frames.
By counting the number of detections per magnitude
bin we can readily quantify the contamination from noise peaks.
It was found that only 7\% of the detections
have $U$ magnitudes brighter than $27$ mag while in the $R$ frame about
9\% have magnitudes brighter than $27$ mag.
With this low level of noise contamination, we decided not to make any
corrections for false detections.

\section{Analysis}
\label{sec:analysis}

The number count method is based on the following property of cluster
lenses.
Under
the standard assumption that the number density of background galaxies
follows a power law,
$n_0(S) \propto S^{-\beta}$ for a given flux limit $S$, one finds a
simple relation between the
magnification of the lens, $\mu$, and $n(\bsy{\theta};>S)$, the number density
of background objects as observed through the lens,
\be{eq:nn0_S}
 {n(\bsy{\theta};>S) \over n_0(S)} = \mu^{\beta - 1}(\bsy{\theta}).
\ee
Here $\bsy{\theta}$ is the position on the sky.
In practice, one observes the local galaxy distribution down to a
limiting magnitude, $m$, so that the cumulative normalized number
density is often written as
\be{eq:nn0_m}
 {n(\bsy{\theta};<m) \over n_0(m)} = \mu^{2.5\alpha - 1}(\bsy{\theta}),
\ee
with $\alpha \equiv \rmd \log N / \rmd m = 0.4 \beta$ being the slope
of the (unlensed) differential galaxy number counts. By inverting
Eq.~(\ref{eq:nn0_m}), it is therefore possible to measure the magnification
field, $\mu(\bsy{\theta})$, provided that a sample of background
sources can be isolated, e.g.~on the basis of color and magnitude
(Sec.~\ref{sec:select}), and that the parameters $\alpha$ and $n_0$
for this sample can be determined to a sufficient accuracy.
In the following sections, we shall discuss these issues, and
present detailed analysis of the observed density field $n/n_0$ in the
form of radially averaged depletion curves and maximum likelihood
analysis.

\subsection{Source extractions}
The first step in the analysis of the final images is to extract
the sources. 
Source extraction was done with both SExtractor
(\cite{Bertin+Arnouts;1996}) and
ImCat (\cite{Kaiser:ImCat}), with comparable results. The ImCat
results were used to identify stars, based on half-light radii, but
the rest of the
analysis below is based on the information extracted by SExtractor.
The full set of SExtractor object catalogs used here is available at
the CDS and ADC catalog archives.

We have performed extensive analysis of simulated data, using noise
characteristics from the final reduced images, to find optimal
adjustments of the various parameters of the detection software (see
\cite{ThomasMScThesis;2000} for details).
For SExtractor, these considerations resulted in choosing a detection
threshold of $1 \sigma$ per pixel,
combined with a Gaussian filtering kernel (FWHM of 5 pixels,
roughly corresponding to our seeing of 6 pixels), and a minimum
detection area of 10 pixels. The complete SExtractor
parameter files used are available at the aforementioned website.

When selecting objects for further analysis from the catalogs produced by
SExtractor, we leave out objects which have internal flags larger than
3. This excludes objects with saturated pixels, objects that have been
truncated, and objects with corrupt isophotal or aperture data.

\subsection{Galaxy number counts}

The original motivation for using $U$ band observations for
depletion studies was the indication of a break in the galaxy number
counts at faint $U$ magnitudes reported by
\cite{Williams+Blacker+Dickinson+;1996}. They found a logarithmic
slope $\alpha \approx 0.4$ for $U_{300} \leq 26$
dropping to $0.05$ in the range $26 < U_{300} < 28$. A more
detailed analysis of the HDF number counts by
\cite{Pozzetti+Madau+;1998} shows that the slope drops to $\alpha = 0.135$ at
$U_{300} = 25.75$, with an allowed range of $0.08 < \alpha <
0.16$.

The differential number counts in our field are shown in
Fig.~\ref{fig:dndm} for the $R$ and $U$ band.
The overall higher level of bright $R$ band objects in
our data relative to the counts from
\cite{Smail+Hogg+Yan+;1995}, \cite{Williams+Blacker+Dickinson+;1996},
and \cite{Hogg+Pahre+;1997}
is simply due to the presence of the cluster. A similar effect is seen
in the $U$ band. The net result of this and the incompleteness at
faint magnitudes is to lower the logarithmic slope of the differential
number counts. This is evident from Fig.~\ref{fig:dndm}a, where we
find a slope of $0.21$ in the magnitude range $20 \leq
$\RAB$ \leq 25$, compared to an estimate of $0.27$ for the other data in
the same magnitude range.

It is evident that our data is incomplete beyond \UAB $\approx
26$ and \RAB $\approx 25.5$. This is consistent with our completeness
estimates, but as shown by \cite{Gray+Ellis+Refregier+;2000}, the
incompleteness does not affect the depletion analysis, as long as the
field used to estimate the average background density has the
same completeness characteristics. It means, however, that we are
unable to determine the slope at the faint end of the luminosity function,
and must rely on estimates from deeper observations.
The slope in the $U$ magnitude interval used in the depletion analysis
(section \ref{sec:ere}) is likely to be varying from
a value close to $0.4$ at \UAB = 24 to the value $0.135$ found by
\cite{Pozzetti+Madau+;1998} at \UAB $\approx 26$.
Similarly, the $R$ band slope is dropping from a value of $\alpha_R =
0.3$ or so at bright magnitudes
(\cite{Smail+Hogg+Yan+;1995,Hogg+Pahre+;1997}) to 0.18 at the faint
end (\RAB=27) of the selected magnitude interval, in accordance with
the slopes found in the $V$ and $I$ bands from HDF data
(\cite{Williams+Blacker+Dickinson+;1996}, see also
\cite{Madau+Pozzetti;2000}).

\mytwofigure{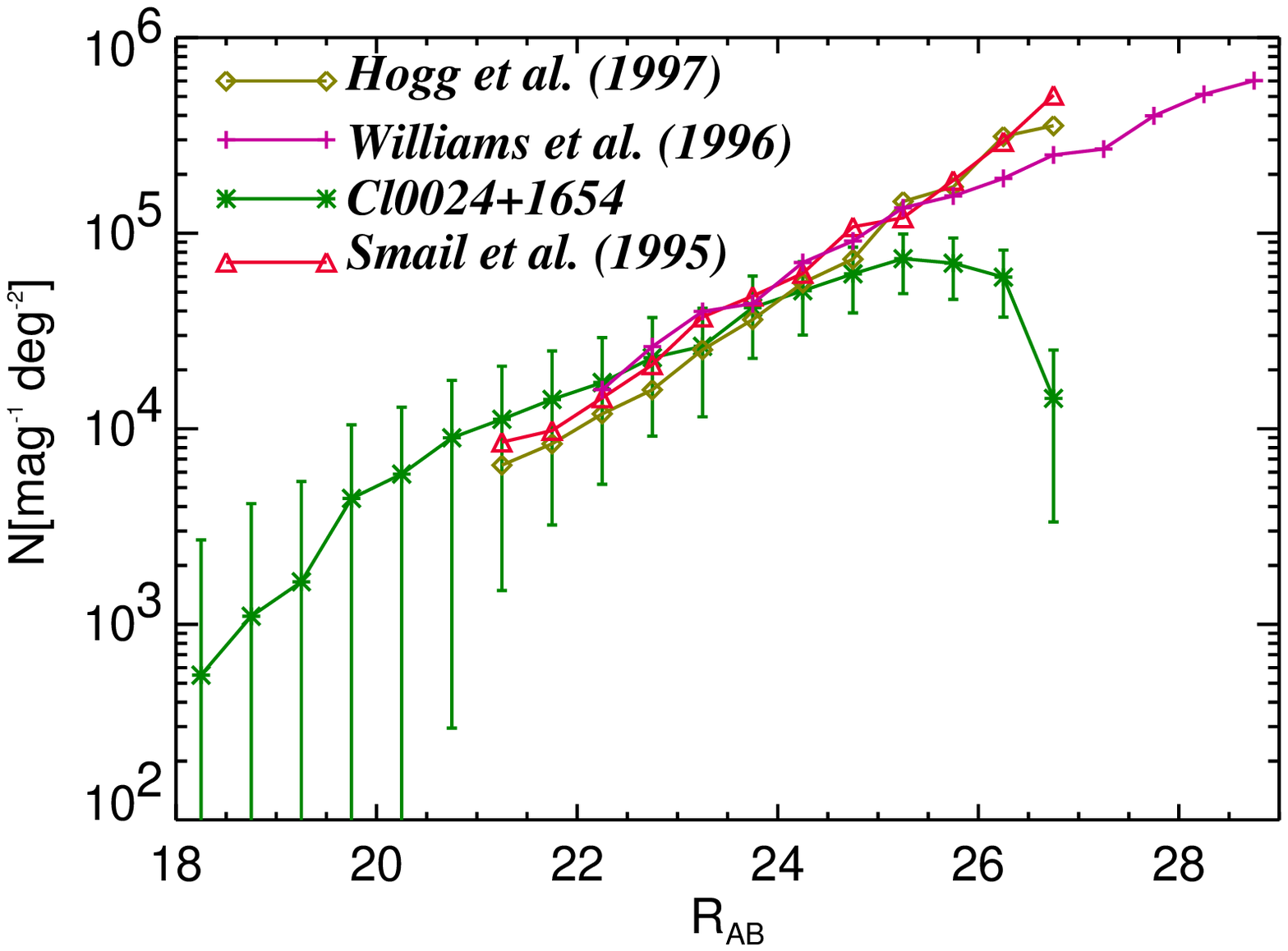}{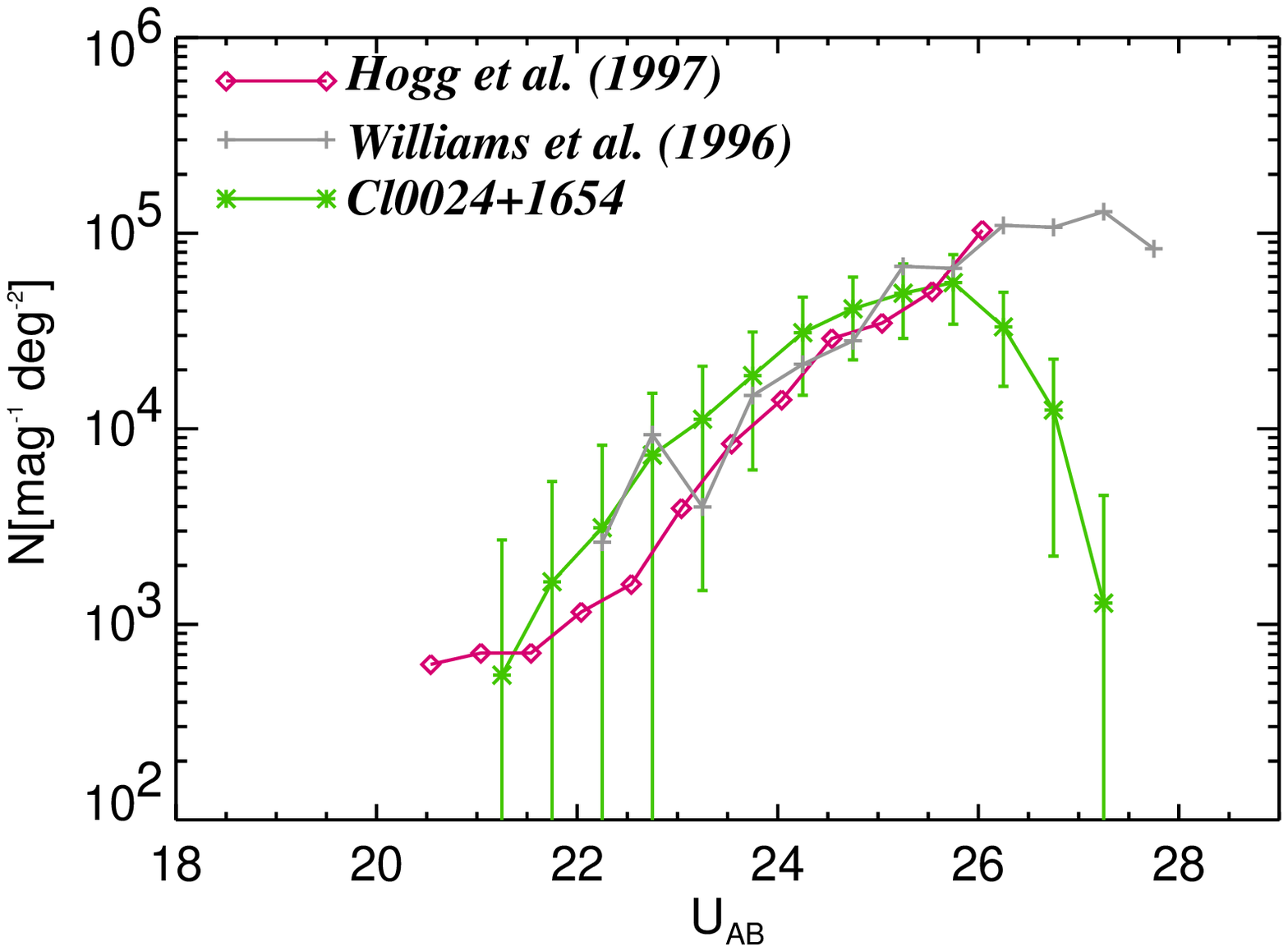}{}{}
{(a) The \RAB~number counts compared to $R$ band data from
\protect{\cite{Hogg+Pahre+;1997}}, and
\protect{\cite{Smail+Hogg+Yan+;1995}},
and $I$ band data from
\protect{\cite{Williams+Blacker+Dickinson+;1996}}.
The effect of the cluster is more pronounced
at bright magnitudes, while incompleteness sets in around \RAB=25.
(b) The \UAB~number counts compared to data from
\protect{\cite{Williams+Blacker+Dickinson+;1996}}, and
\protect{\cite{Hogg+Pahre+;1997}}. Again, the presence of the cluster
results in overall higher counts, and the turn-over at faint
magnitudes reflects the incompleteness beyond \UAB=26.
}
{fig:dndm}

\subsection{Color-magnitude diagrams}
\label{sec:cmd}
We find the \URAB~color by first detecting and measuring the magnitudes
in the $U$ band.
The $U$ catalog is then matched to the $R$ band image by running SExtractor
in ``assoc'' mode with a search radius of 6 pixels,
appropriate to compensate for our seeing and error in the alignment
between the $U$
and $R$ images. In this way, 1229 objects are found when associating
the $U$ image with the full $R$ catalog.
These objects comprise our ``matched'' catalog.

In Figs.~\ref{fig:U-R}a and \ref{fig:U-R}b we show the
color-magnitude diagram for all the $1169$ objects in the matched catalog
after removal of $60$ stars (the distinction between stars and
galaxies was based on the analysis
of half-light radii). The formal detection limits in $U$ and $R$ are
indicated by dashed lines, and the color and magnitude cuts used to
select background objects are shown with dash-dotted lines (see
Sec.~\ref{sec:select}). The bluing trend at faint $R$ and $U$ magnitudes
(e.g.~\cite{Hogg+Pahre+;1997}) is reflected in the negative slope of
the lower edge of the ``wedge'' shaped region that the objects
populate. This effect is especially pronounced in Fig.~\ref{fig:U-R}a.

\mytwofigure{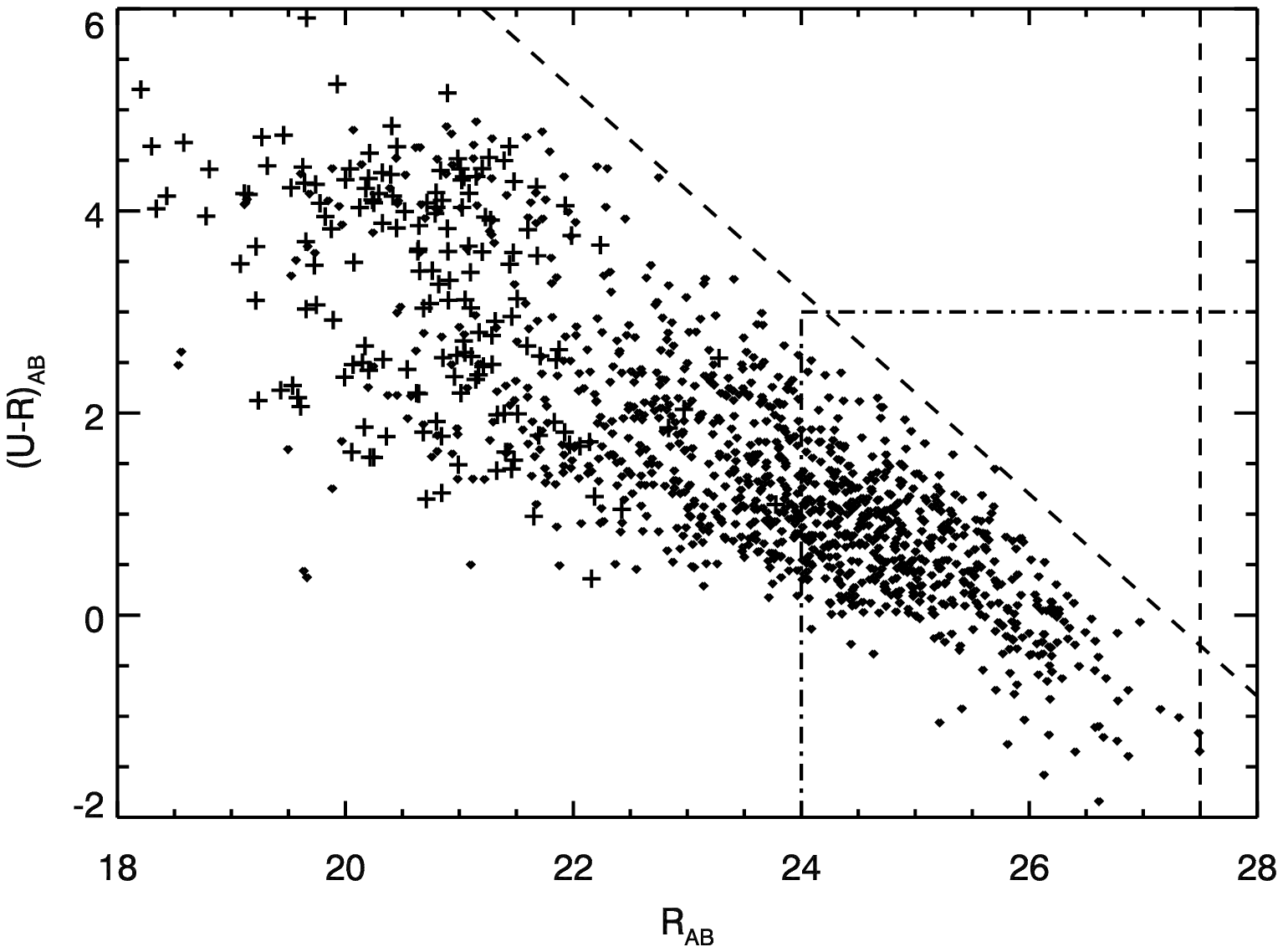}{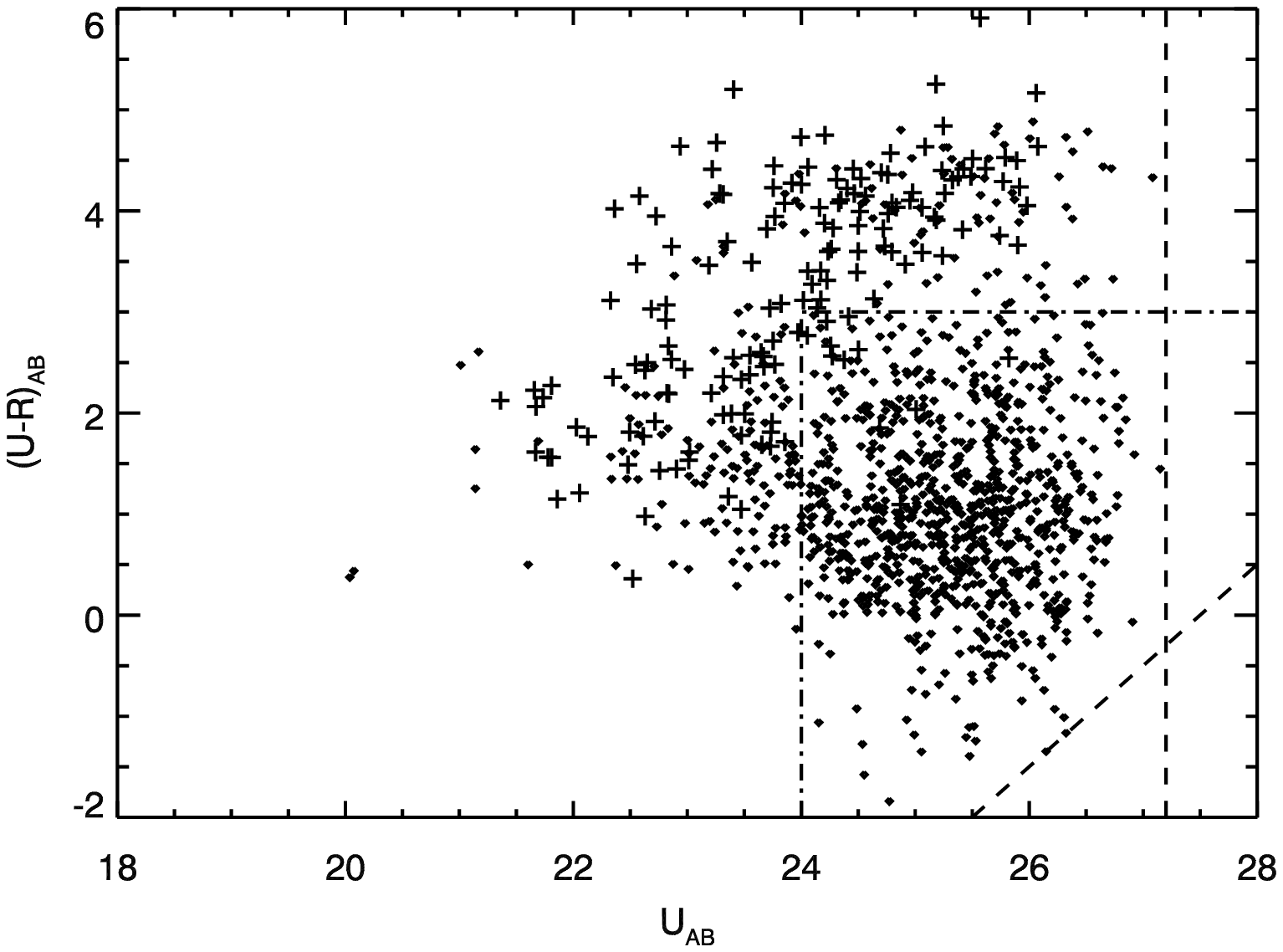}{}{}
{
(a) The \URAB~vs~\RAB~color-magnitude diagram. The formal detecion
$3\sigma$ limits, \UABlim = 27.2 mag and \RABlim = 27.5 mag, are indicated with
dashed lines.
Dash-dotted lines indicate the color (\URAB $\leq 3$) and
magnitude (\RAB $\geq 24$) limits used
to mask out foreground objects in the $R$ band.
The upper edge of the ``wedge'' shape is defined by the
limiting magnitude in $U$, while the slope of the lower edge clearly
reflects a bluing trend towards fainter magnitudes. Note also the
distinct low density around \URAB=3. Confirmed cluster galaxies
are marked by '+'.
(b) The \URAB~vs~\UAB~color-magnitude diagram.
Dash-dotted lines show the color (\URAB $\leq 3$) and
magnitude (\UAB $\geq 24$) limits used to mask out foreground
objects in the $U$ band. The general features are the same as in (a)
}
{fig:U-R}

The color-magnitude diagrams clearly show a lower density of objects
around \URAB=3, also seen in $U$ and $R$ band data we have obtained
for the cluster MS1621+2640 (van Kampen et al., in preparation).
By combining our photometric data with redshifts from
the survey of \cite{Czoske+Soucail+Kneib+;1999} and \cite{Dressler+;1999}
for the
CL0024+1654 field, we find that the \URAB~color of the cluster
galaxies shows a distinct bi-modal distribution, with a
minimum around \URAB=3 and a large fraction above \URAB=3. This
distinction is not seen for the field galaxies in the redshift survey,
suggesting that mostly cluster galaxies will be
discarded by selecting only galaxies bluer than \URAB=3.
This can also be inferred from the distribution of confirmed cluster 
galaxies, marked by '+', in the color-magnitude diagrams.
The radial gradient observed for $g-r$ color in many clusters (see
e.g.~\cite{Morris+Hutchings+Carlberg+;1998}) is also seen in the
\URAB~color. These features are suggestive of a strong
correlation between star formation activity in cluster galaxies and the cluster
environment. We shall investigate this aspect of the data in
greater detail in a forthcoming paper (van Kampen et al., in preparation).

\subsection{Selection of background candidates}
\label{sec:select}
For the depletion analysis, it is important to remove cluster and foreground galaxies from the
object catalogs as the number density of cluster galaxies
typically increases towards the center, interfering with radial
gradients in the number density of the background.
In line with the discussion above on color-magnitude diagrams,
we chose to use simple limits on magnitudes and color, restricting the
depletion analysis to the magnitude ranges $24 \leq (R,U)_{\mrm{AB}}
\leq 27$ and to objects bluer than $(U-R)_{\mathrm{AB}} = 3$.
The limiting $U$ band
magnitude of 27.2 implies that $R$ band objects in the
magnitude range $24 \leq R_{\mrm{AB}} \leq 27$ that have a
well defined counterpart in the $U$ band image are all bluer
than \URAB=3, as can be seen from figure \ref{fig:U-R}a.
Therefore, the exclusion of objects redder than \URAB=3 will
only affect the selection of background objects in the $U$ band.

The background samples are constructed by first selecting
objects in the magnitude ranges $24 \leq (R,U)_{\mrm{AB}}
\leq 27$ from the full $R$ and $U$ band catalogs. In order
to exclude the red (cluster) objects, one could in principle extract
the background sample from the matched catalog, with the
additional constraint on color. Doing this, however,
one tends to miss
primarily the faintest objects, partly because some of the $U$
objects are simply too faint in $R$ to be detected, and vice versa, as
well as because the association is more difficult the fainter the
object is. We avoid this drawback by constructing masks
(using the shape parameters estimated by SExtractor)
covering the $(U-R)_{\mathrm{AB}} \geq 3$ objects in the
matched catalog. The objects that fall on these masks are
then deleted from the background samples.

In addition, the masks are augmented with the
bright objects ($(R,U)_{\mrm{AB}} < 24$), to
cover the parts obscured by the foreground
and cluster candidates in the corresponding image (see figure \ref{fig:mask}).
The estimated (Poisson) errors for the depletion curves
are corrected for
the effective area of each annulus of the radial averaging
by subtracting off the number of pixels in the mask from the
pixelized area of the annulus. The correction due to the mask is quite
important for the innermost bins, especially in the $R$ band, where
the bright cluster galaxies obscure a large fraction of the
area. Hence, the innermost bins have the highest uncertainty due to
the small effective area.

\mytwofigure{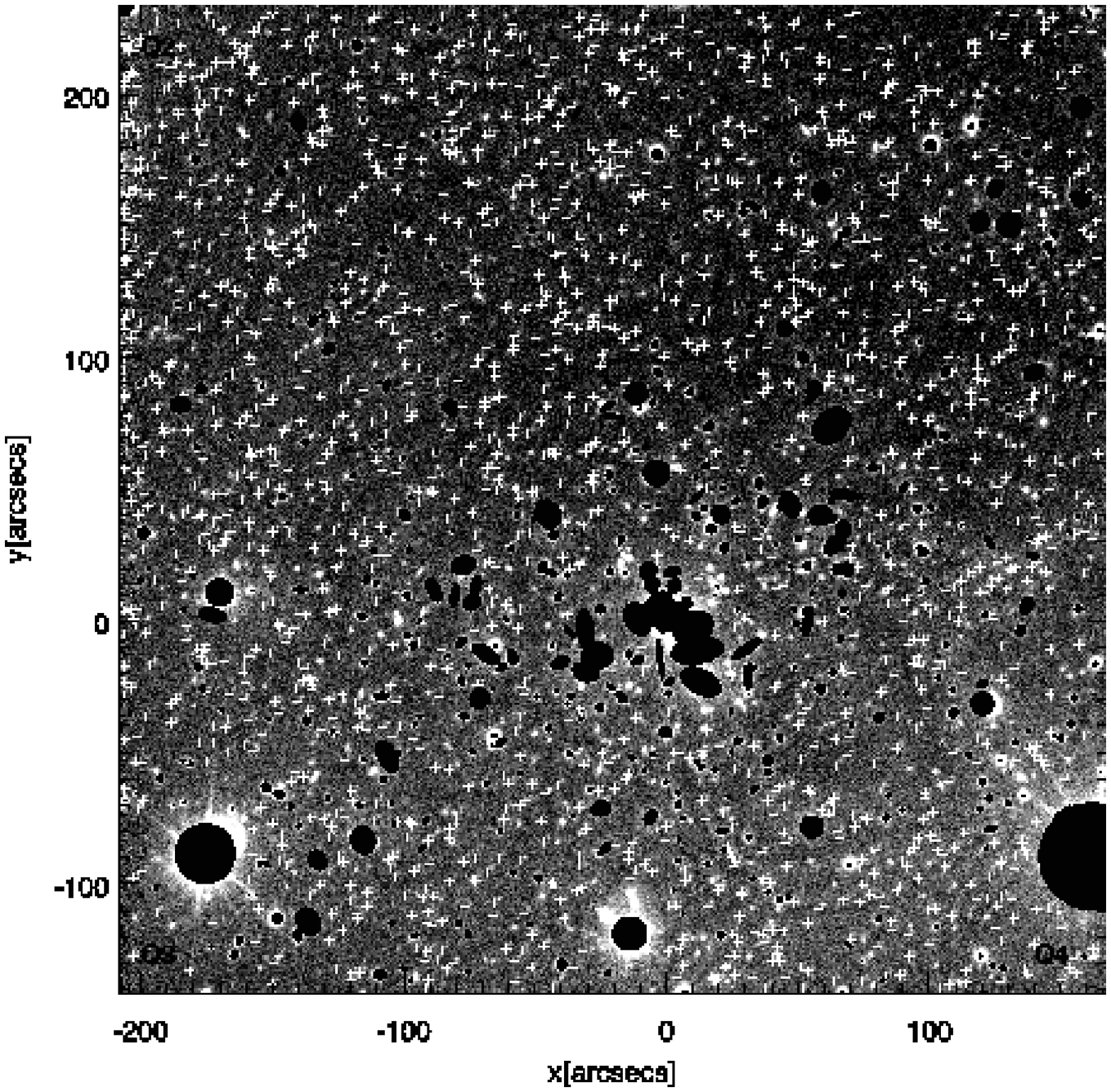}{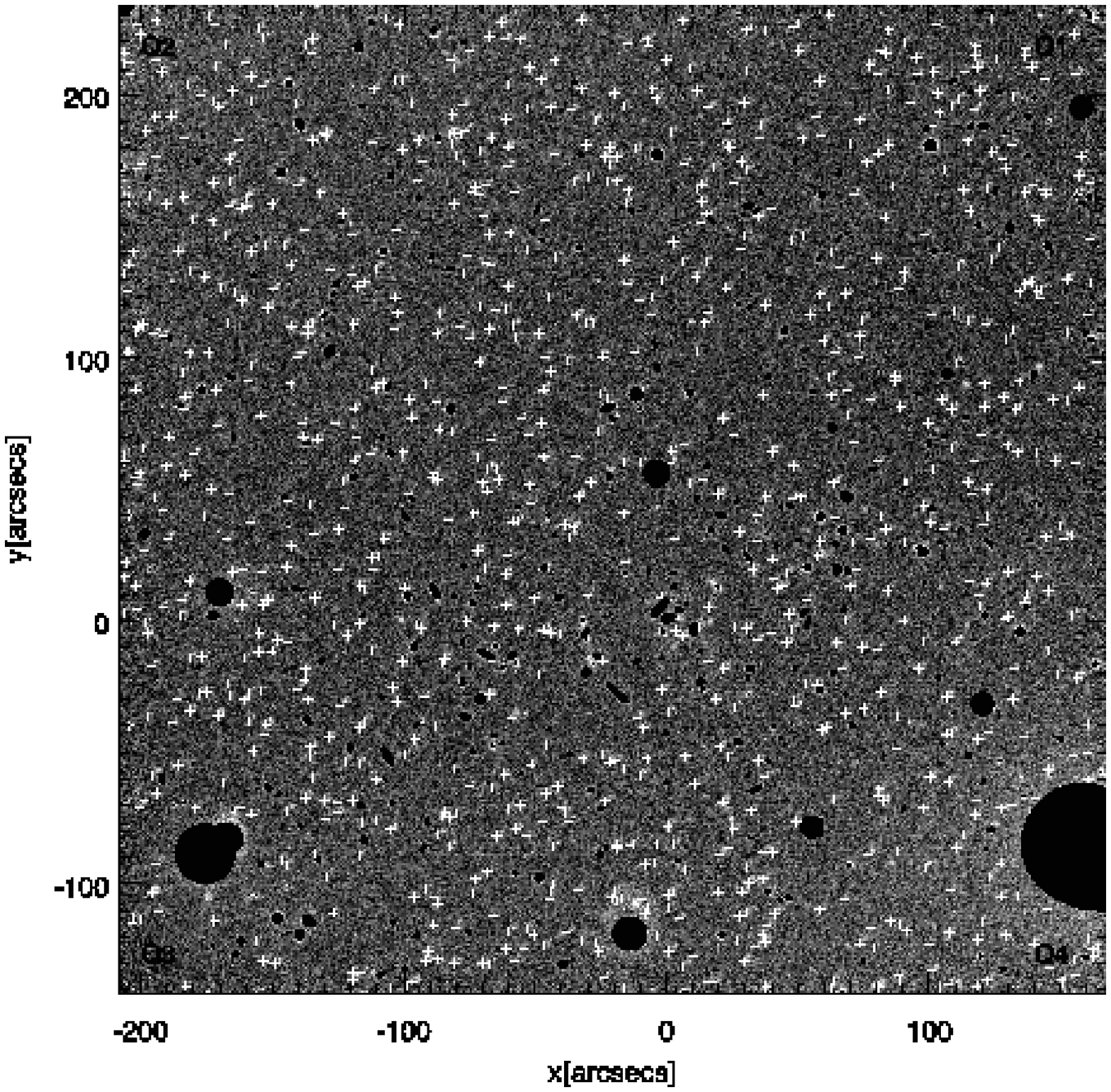}{}{}
{
(a) The $R$ band mask used in the depletion analysis superimposed
on the $R$ band image of CL0024+1654. Objects
brighter than \RAB $= 24$ and redder than \URAB $= 3$ are masked out.
The background objects are marked with `+'.
(b) Same as (a) for the $U$ band.
Objects
brighter than \UAB $= 24$ and redder than \URAB $= 3$ are masked out.
}
{fig:mask}

\subsection{Depletion analysis}

Radial averaging is typically used to boost the
significance of the depletion signal
(e.g.~\cite{Taylor+Dye+Broadhurst+;1998,Mayen+Soucail;2000}). The
number density is estimated by counting the number of objects in
annuli around the cluster center. This average density is normalized by
an (independent) estimate of
the unlensed background density of the same class of objects
resulting in a radial depletion
curve. The width of the annuli, $\Delta r$, may be adjusted to obtain a
reasonable balance between smoothing and shot noise.
We calculate the number density averaged over $10\arcsec$ wide annuli
as a function of radius from the
cluster center, taken to be the center of light
($\alpha(\mrm{J2000}) = 00^{\mrm{h}}26^{\mrm{m}}35.8^{\mrm{s}}$ and
$\delta(\mrm{J2000}) = +17^{\circ}09^\prime42.3\arcsec$) of the $R$
image within $100\arcsec$ radius after a few bright stars and
a foreground galaxy have been masked out.
The number counts are normalized to the background density, ${n}_0$,
estimated as the average density of the corresponding objects in the
annulus between $150'' < r < 240''$. This assumes that the cluster
does not affect the background significantly beyond
$150\arcsec$. (Ideally, the unlensed background density should be
estimated from a separate field, known not to be affected by strong
foreground lenses.)
We find values of $\bar{n}_0(U) = 28 \pm 2$
arcmin$^{-2}$ and $\bar{n}_0(R) = 49 \pm 7$ arcmin$^{-2}$, with 1084 and 1780
selected background objects in the $U$ and $R$ band, respectively. (The
uncertainties quoted are jackknife estimates, see below.)

The obscuration by the masks
(separate for each band) is taken into account in determining an
effective area, $A_{\mrm{eff}}(r_i)$, for each annulus, $r_i - \Delta
r/2 \leq r < r_i + \Delta r/2$. The uncertainties may as a first
approximation be
taken to be due only to Poisson noise in the counts, so that
\[
  \Delta N(r_i) = N_{\mrm{exp}}(r_i)^{1/2}
                = (n_0 A_{\mrm{eff}}(r_i))^{1/2}
\]
where $N(r_i)$ are the actual number counts in the annulus centered on
$r_i$, and $N_{\mrm{exp}}(r_i) = {n}_0 A_{\mrm{eff}}(r_i)$ is the
expected number of counts in the annulus. Then,
\[
  {\Delta n(r_i) \over n_0} = (n_0 A_{\mrm{eff}}(r_i))^{-1/2}.
\]
Given an uncertainty in the absolute background density, $\Delta n_0$,
a more conservative estimate of the uncertainty is
\be{eq:dnn0_2}
  \Delta \left( { n(r_i) \over n_0} \right) = \left( 
                         \left({\Delta n(r_i) \over n_0} \right)^{2}
                      +  \left({n(r_i) \Delta n_0 \over n_0^2}\right)^{2}
                         \right)^{1/2}
.
\ee
We have made simple estimates of the intrinsic uncertainty in ${n}_0$ due to
non-uniformity of the background, effects of bright objects, etc,
by masking out each of the four quadrants of the images in turn,
numbered counter-clockwise from the NE quadrant.
The four values of the background obtained in this way are then used
to form a standard jackknife estimate the variance of $n_0$.
We find that $\sigma_U(n_0) = 2.2$ and $\sigma_R(n_0) = 7.2$.
The errors indicated on Fig.~\ref{fig:nn0}a and \ref{fig:nn0}b include a
fractional uncertainty $\eta = \Delta n_0/n_0$ of
$\eta_R = 7.2/49 = 0.15$ and $\eta_U = 2.2/28 = 0.08$,
respectively. Note, that an error in ${n}_0$ only affects the
normalization of the depletion curve, while the relative
depletion does not change.
The estimated background density in the $R$ band is strongly
influenced by quadrant 2, where the density of selected objects is
markedly higher than in the other three quadrants. The form of the
depletion curves is however robust. The variation in the $R$
band may at least partly be explained by a higher background flux in quadrants
3 and 4, due to bright stars. The background flux in the $U$ band is more
uniform across the whole field.
The estimated values of $\eta$ should only be taken as an indication of the
level of uncertainty in $n_0$.

The depletion curves in the $R$ and $U$ band are shown in
Fig.~\ref{fig:nn0}. A clear sign
of depletion is seen in both bands. In the $R$ band the density falls steadily
from a radius of around $100 \arcsec$ all the way to the center. Most
lens models predict a turn-up in the density inside the innermost
critical line; the indication seen here in the innermost two bins is
non-significant due to the obscuration of the foreground
galaxies. On the other hand, it is easier to detect faint background
objects closer to the cluster center in the $U$ band, since the
dominant cluster galaxies are all quite red.
Given that we are successful in identifying
cluster members, the observed turn-up is quite sharp and significant in the
$U$ band.

Our data is not complete to sufficiently faint magnitudes to
allow estimation of the slopes $\alpha_R$ and $\alpha_U$, for the magnitude ranges
involved. However, we
find that most of the depletion signal in the $U$ band is due
to galaxies with $24 < U_{\mrm{AB}} < 26.5$, which indicates that the slope
has already dropped well below $0.4$ in this magnitude range. This is
at odds with the results of \cite{Hogg+Pahre+;1997}, who found a slope
of $0.467$ down to $U_{13} = 25.5$, roughly corresponding to \UAB=26.3.
The recent investigation by \cite{Volonteri+Saracco+;2000} of the
HDF-S shows no evidence of a break either, down to the faintest
magnitude bin of \UAB=26.5.

Although the
data may not warrant a detailed quantitative interpretation of the
depletion curves, especially when taking the uncertainty
in the slope, $\alpha$, and unlensed background estimate, $n_0$,
into account, it is in principle possible to use multiband depletion
measurements to constrain the properties of the background
population (see
e.g.~\cite{Taylor+Dye+Broadhurst+;1998,Bartelmann+Schneider;2000,Mayen+Soucail;2000},
and references therein) due to the dependence on the slope of the
luminosity function. The redshift distribution of the background
objects will also affect the width and position of the minimum of the
depletion curve, essentially through the dependence of the Einstein
radius on the lens-source distance. In Fig.~\ref{fig:nn0}, we compare
the radially averaged data to the singular isothermal sphere (SIS)
model described in section \ref{sec:ere}.
In case of the $R$ band depletion curve,
results for $\alpha_R = 0.15$ and $0.2$ are shown, with the Einstein
radius $\rE=25\arcsec$.
In addition, we show the theoretical
depletion curve for $\rE=40\arcsec$ and $\alpha_R=0.32$. The $U$ band
depletion is compared to models with $\alpha_U=0.15$ and
$0.2$, with a fixed Einstein radius of $\rE = 20\arcsec$, as well as
to $\alpha_U = 0.32$ and $\rE=32\arcsec$. (This choice
of parameters is dictated by results of the
maximum likelihood analysis in section \ref{sec:ere}.) We note that
the depletion curve is
better represented by the $\rE=25\arcsec$ and $\rE=20\arcsec$
models in the $R$ and $U$ bands, respectively.

\mytwofigure{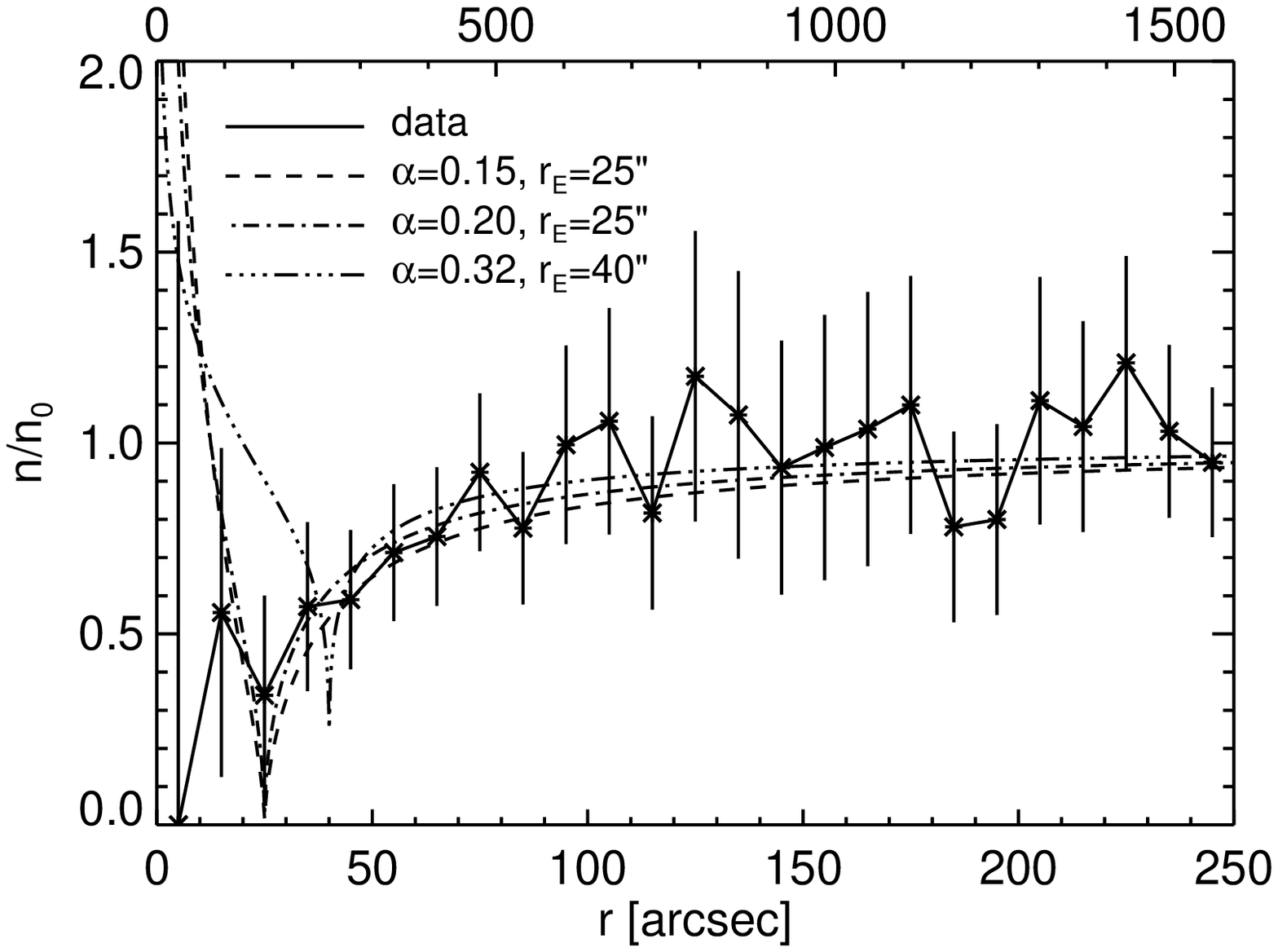}{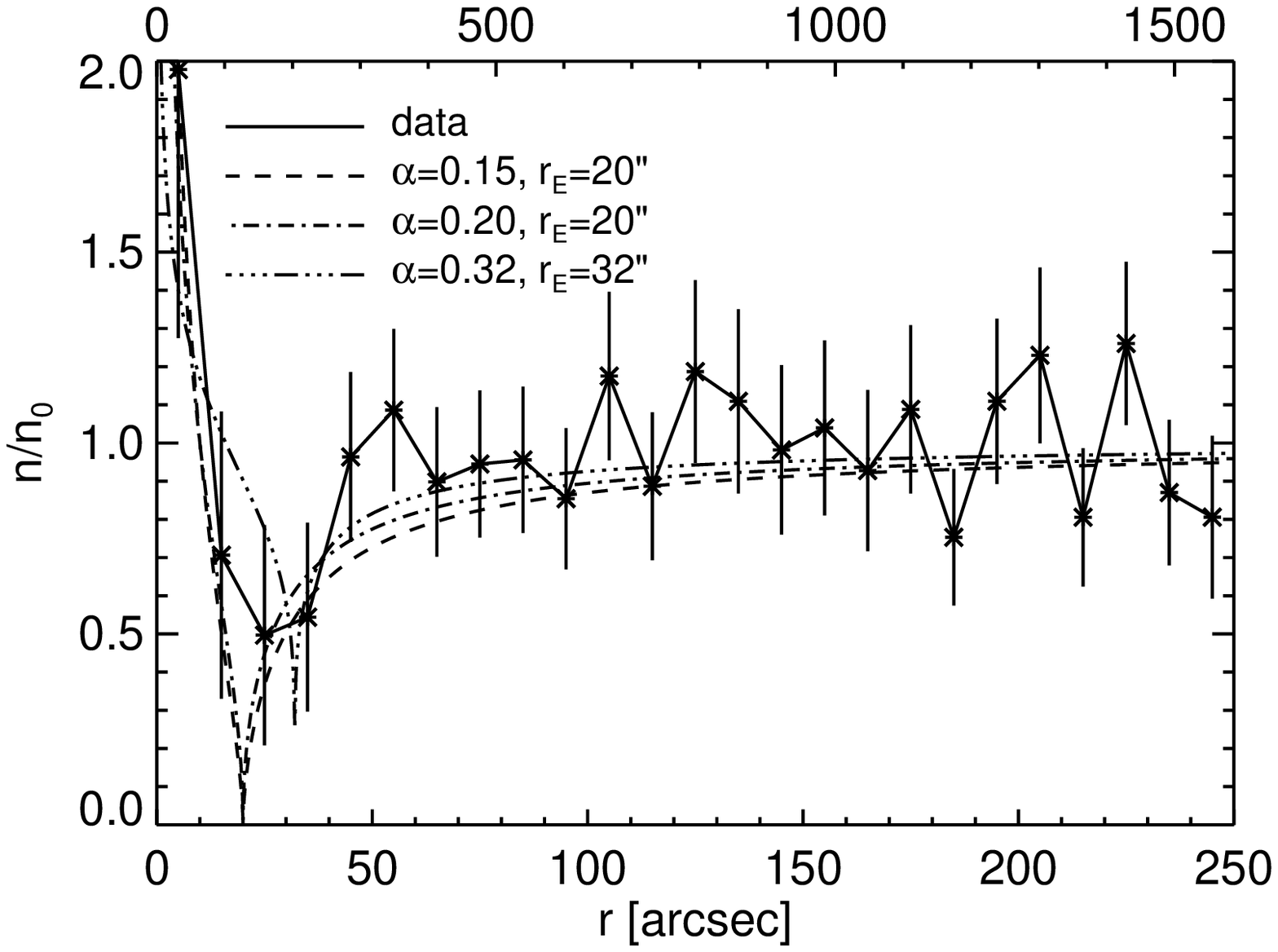}{}{}
{
(a) Normalized density of faint $R$ band galaxies as a function of radius
from the center of CL0024+1654. The selected magnitude range is $24
\leq R_{\mrm{AB}} \leq 27$, and the radial binsize is $10\arcsec$.
The upper radial
scale shows the distance from the center in $h_{50}^{-1}$ kpc, assuming an
$\Omega=1$ cosmology. The data is compared to three singular isothermal
sphere (SIS) models, with \{$\rE$,$\alpha$\}=\{$25\arcsec$, 0.15\},
\{$25\arcsec$,0.2\}, and \{$40\arcsec$,0.32\}.
(b) Same figure for the $U$ band. The selected magnitude range is $24
\leq U_{\mrm{AB}} \leq 27$. Here, the SIS models shown have
\{$\rE$,$\alpha$\}=\{$20\arcsec$, 0.15\},
\{$20\arcsec$,0.2\}, and \{$32\arcsec$,0.32\}. 
}
{fig:nn0}

\subsection{Einstein radius estimates}
\label{sec:ere}

We use the maximum likelihood approach developed by
\cite{Schneider+King+Erben;2000} and \cite{Gray+Ellis+Refregier+;2000}
to investigate the depletion in more detail. 
A singular isothermal sphere (SIS) model, where the magnification is a
simple function of the distance from the center of the lens, $r$, and
its Einstein radius,
\[
\mu_{\mrm{SIS}}(r;r_{\mrm{E}}) = \left| {1 \over 1-r_{\mrm{E}}/r} \right| ,
\]
is used to calculate the likelihood as a function of a single
parameter, the Einstein radius $r_{\mrm{E}}$.
Recall, that the theoretical depletion curve is in this case given by
\[
 {n(r) \over n_0} =  \left| {1 \over 1-r_{\mrm{E}}/r} \right|^ {\beta - 1}
\]
where the slope of the luminosity function is $\beta = 2.5\alpha$. 
For simplicity, we adopt a constant slope, noting that a proper
treatment of our data would take the variation in $\alpha$ into
account, since we are very likely
probing a transition region in both the $R$ and $U$ bands. We take the
(unknown) uncertainty in the unlensed background density into
account, using the likelihood function (see
\cite{Gray+Ellis+Refregier+;2000}, Eq.~(10)),
\bea{eq:likely_1}
  \log L_\mu^{(\eta)} \equiv l_\mu^{(\eta)} & = & -n_0 I + (\beta - 1) \sum_{i=1}^{N} \log
\mu(\bsy{\theta}_i) \nonumber \\
  & &  + N \log n_0 - { (n_0 - \bar{n}_0)^2  \over 2(\eta \bar{n}_0)^2  }
\eea
where $I = \int \rmd^2\bsy{\theta}
(\mu(\bsy{\theta}))^{\beta-1}$, and
\be{eq:likely_2}
  { n_0 \over \bar{n}_0 } = \nu + (\nu^2 + \eta^2 N)^{1/2}
\ee
with $\nu = {1 \over 2} (1-\eta\bar{n}_0 I)$. The first term in
Eq.~(\ref{eq:likely_1}) stems from the likelihood of observing a
total of $N$ galaxies in the whole field, while the second term is the
log-likelihood for finding those $N$ galaxies at the observed positions
$\bsy{\theta}_i = (r_i,\theta_i)$ in polar coordinates. The third term
is just a
normalization, but is needed in this case as we include prior
information on $n_0$, taken to be a Gaussian with mean $\bar{n}_0$ and
dispersion $\eta \bar{n}_0$. The maximum likelihood estimate of
$n_0$ is then given by Eq.~(\ref{eq:likely_2}), and is inserted back
into Eq.~(\ref{eq:likely_1}) to obtain the log-likelihood. Note, that the
model parameters ($\rE$ in the SIS case) are all hidden in the
magnification $\mu$.

To assess the effect of
the uncertainty in $n_0$ we study four values of the dispersion
by setting $\eta = 0$, 0.05, 0.1, and 0.15. The highest
value is probably a conservative estimate taking clustering of the
background into account, given the latest measurements of the two point
correlation function (\cite{Fynbo+Freudling+Moller;2000}), which
indicate insignificant correlation beyond $10\arcsec$ for the
magnitude range considered here. The estimated amplitude
of the two point correlation function in the $U$ band is even lower
(\cite{Brunner+Szalay+Connolly;2000}),
making the $U$ band depletion less sensitive to errors due to clustering.
Possible bias in the estimated
unlensed density is studied by shifting $\bar{n}_0$ up by 10\%. We
would expect the estimated $\bar{n}_0$ to be biased low relative to
the true value $n_0$, since the depletion due to the cluster probably
extends beyond the inner radius $r_{\mrm{i}}=150\arcsec$. Contamination by cluster members
would however work in the other direction, and it is difficult to
disentangle the two effects with our data. By varying the inner radius
$r_{\mrm{i}}$ from $150\arcsec$ to $220\arcsec$ we find that the $U$ band
background density is quite stable, while in the $R$ band $\bar{n}_0$
does indeed increase by roughly 10\%, going to 54 arcmin$^{-2}$ at
$r_{\mrm{i}}=200\arcsec$.

We also study the dependence of the estimated Einstein radius on the
slope in the range $0.1 \leq \alpha \leq 0.4$.
In the $U$ band, $\alpha_U
= 0.135$ is appropriate for \UAB $> 25.5$, but since 50\% of the
selected background galaxies
in the magnitude range $24 \leq $\UAB$ \leq 27$ are brighter than
this, we expect the effective slope to be steeper.
Similarly, in the $R$ band, the slope at the bright end is of order
0.32, dropping to $\approx 0.18$ at the faint end (\cite{Smail+Hogg+Yan+;1995,Williams+Blacker+Dickinson+;1996,Madau+Pozzetti;2000}). As
50\% of the background candidates
have \RAB $> 25.3$, the effective slope probably lies in this range.
The uncertainties, $\Delta r_{\mrm{E}}$, quoted below are estimated as
the width of the likelihood function, $L_{\mu}^{(\eta)}$,
\be{eq:delta_rE}
\Delta \rE = { \int (\rEp - \rE)^2 L_{\mu}^{(\eta)} \rmd \rEp
         \over \int L_{\mu}^{(\eta)} \rmd \rEp },
\ee
where $\rE$ is the estimated value of the Einstein radius, corresponding to
the maximum of $L_{\mu}^{(\eta)}$.

The results of this analysis for the $R$ and $U$ band are shown in
Fig.~\ref{fig:logLrE}, where the estimated Einstein radius, $\rE$, is plotted
as a function of the slope $\alpha$ for values of the dispersion
parameter $\eta$ in the expected range $0-0.15$. The error bars indicate
$\pm \Delta \rE$. The estimated $\rE$
grows with increasing $\alpha$, up to $\alpha=0.4$ where the
theoretical depletion signal vanishes. The effect of increasing
$\alpha$ is to lower the log-likelihood at small $\rE$, so that
local maxima at higher $\rE$ take over as global maxima, causing
discrete jumps in the $(\alpha,\rE)$ plots. Increasing $\eta$ tilts
the $l_\mu^{(\eta)}$ curve upwards at high $\rE$, but for
$\eta \geq 0.05$ the estimated $\rE$ is not very sensitive to this
parameter.
Assuming that our unlensed background
estimate $\bar{n}_0$ is biased low relative to the true value by about
10\%, we find that even the $\eta=0$ estimates of $\rE$ are compatible
with the $\eta \geq 0.05$ cases.
The dependence of the log-likelihood $l_{\mu}^{(\eta)}$ on the
parameters $\alpha$ and $\eta$ is shown in
Fig~\ref{fig:logL} where we plot
$l_{\mu}^{(\eta)}$ as a function of the Einstein radius, $\rE$, for
selected values of these parameters.
We simply tabulate $l_{\mu}^{(\eta)}$ with $\Delta \rE = 0.5\arcsec$
and take the estimated Einstein radius to be the $\rE$ where the tabulated
function is maximized. The dips in the $l_{\mu}^{(\eta)}$ curves
are caused by galaxies that happen to lie close to the current $\rE$
(the critical line being tested), where
the second term in Eq.~\ref{eq:likely_1} dominates, and the probability
of finding a galaxy vanishes.

\mytwofigure{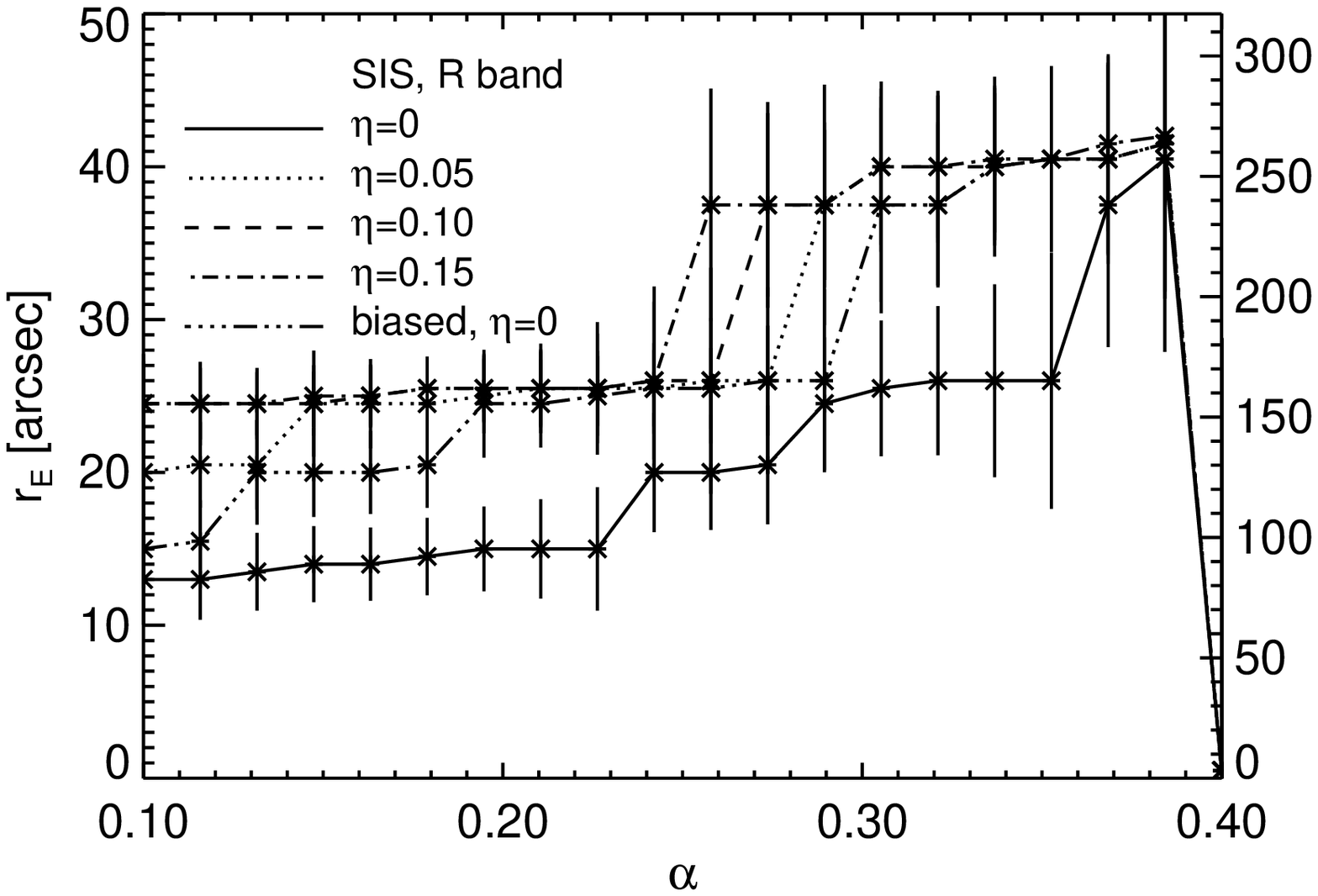}{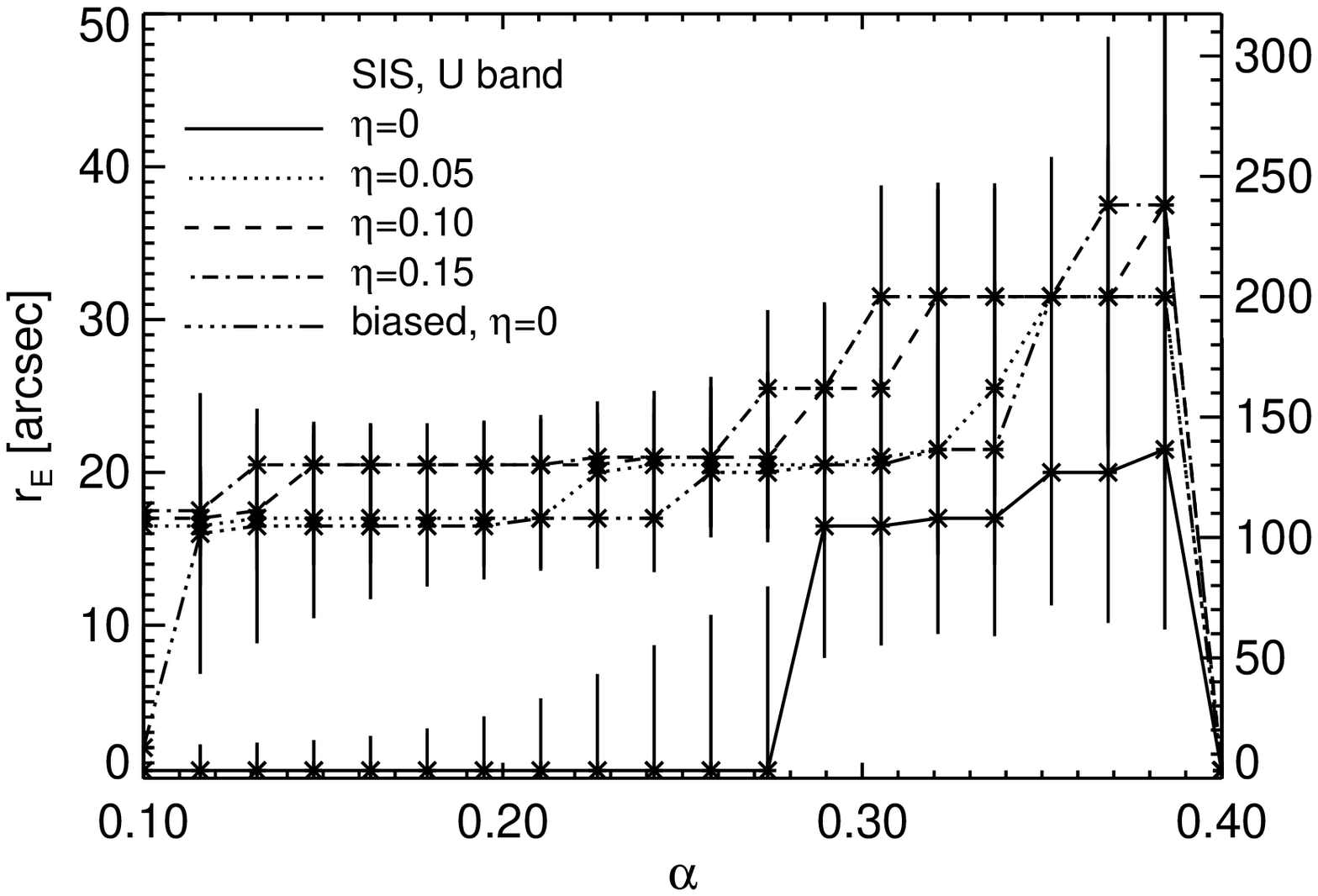}{}{}
{
(a) The estimated Einstein radius in the SIS
model as a function of assumed slope $\alpha$ for the four values of
the dispersion parameter $\eta=0$, 0.05, 0.1, and 0.15, using the $R$
band depletion. The biased model has $n_0 = 1.1 \bar{n}_0$ and $\eta=0$.
The right scale is in $h_{50}^{-1}$ kpc, assuming an $\Omega=1$
cosmology.
(b) Same as (a) using the $U$ band depletion.}
{fig:logLrE}

\mytwofigure{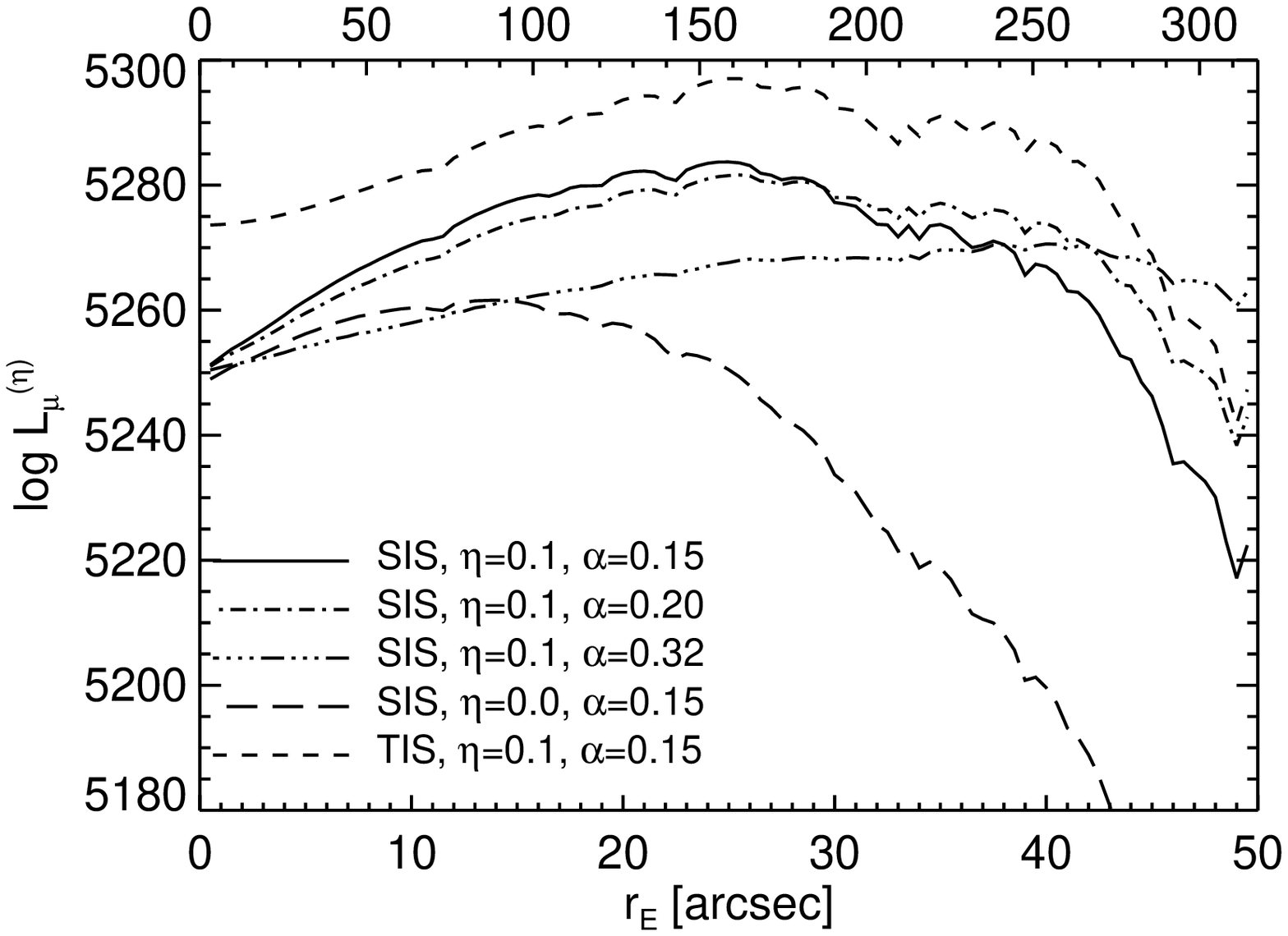}{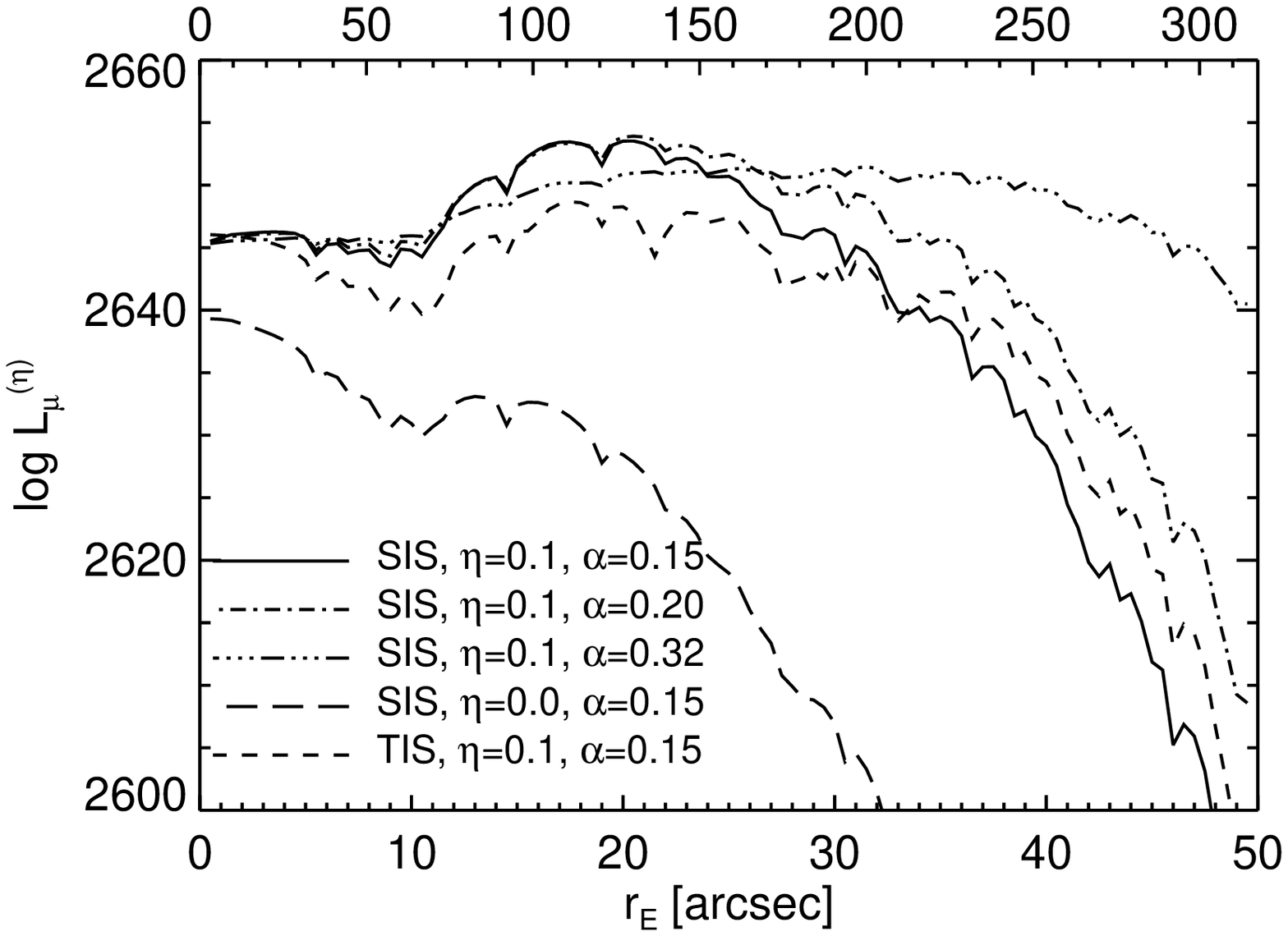}{}{}
{
(a) Maximum likelihood estimation of the Einstein radius in the SIS
model using the $R$
band depletion. The log-likelihood is plotted for the same
parameters as the theoretical depletion curves in
Fig.~\protect{\ref{fig:nn0}}a, i.e.~$\eta=0.1$ and $\alpha=0.15$, 0.2,
and 0.32. In addition, the log-likelihood for the TIS model (see text)
with $\eta=0.1$ and $\alpha=0.15$ is shown. The upper radial
scale shows the Einstein radius in $h_{50}^{-1}$ kpc, assuming an
$\Omega=1$ cosmology.
(b) Same as (a) using the $U$ band depletion.}
{fig:logL}

From Fig~\ref{fig:logLrE}a we see that 
for slopes $\alpha \leq 0.25$
the estimated Einstein radius is $25\arcsec \pm 3\arcsec$ in the $R$
band with a limited range of parameters suggesting a lower value of
$20\arcsec$. The unbiased $\eta=0$ case results in yet a lower value
of around $13\arcsec$. For the rest of the $\alpha$ range, the maximum of
$l_{\mu}^{(\eta)}$ occurs around $40\arcsec$, again with the unbiased
$\eta=0$ case as an exception. Similar features are seen in the $U$
band (Fig.~\ref{fig:logLrE}b), except that for any given set of
parameters the $\rE$ estimates
are always lower than in the $R$ band. In any case, the $U$ band
Einstein radius of $20\arcsec \pm 3\arcsec$ found for $\alpha \leq
0.25$ and $0.1 \leq \eta \leq 0.15$ is within the uncertainty of the
other estimate of $17\arcsec \pm 3\arcsec$ found for the $\eta = 0.05$
and the biased $\eta = 0$ models. The unbiased $\eta=0$ model is
compatible with $\rE=0$ for $\alpha < 0.28$ where the $\rE \approx
17\arcsec$ peak in $l_{\mu}^{(\eta)}$ takes over.
Since the giant arcs seen at around $30\arcsec$ distance from the
cluster center are the multiple images of a galaxy at $z=1.675$
(\cite{Broadhurst+Huang+Frye+;2000}) it seems reasonable to reject
any estimates that exceed this value. It must however be emphasized
that this analysis is based on a single underlying model (SIS) and the
final constraints on the parameters $\alpha$, $\bar{n}_0$, and $\eta$
must come from reliable deep galaxy counts.
To summarize, our analysis indicates that the Einstein
radius may be slightly lower in the $U$ band compared to the $R$ band,
suggesting a lower median redshift for the $U$ band objects.

For comparison, we have also calculated the likelihood function for a
truncated
isothermal sphere (TIS) model, with a core radius of $r_{\mrm{c}} =
10\arcsec$ in accordance with results found from strong lensing
(\cite{Tyson+Kochanski+DellAntonio;1998,Broadhurst+Huang+Frye+;2000}) and
X-ray observations (\cite{Bohringer+Soucail+Mellier+;2000}). The TIS model
is given by equation (8) in \cite{Mayen+Soucail;2000}.
The estimated Einstein radius in the TIS model is almost identical to
the SIS results, as expected from the small core radius. Finally, we
have checked that our results are insensitive to a more stringent
selection of $R$ band objects by redoing the analysis for the
magnitude range $25.3 \leq $\RAB$\leq 27$. For $\alpha \leq 0.2$ and
$\eta \geq 0.05$ this shifts the estimated Einstein radius slightly
outwards, to around $27\arcsec$. This is to be expected, as the
fainter sample will have a higher median redshift.

\section{Discussion and conclusions}
\label{sec:discuss+conclude}
The applicability of lensing methods to determine cluster masses and
density profiles has been a major issue since it became viable to
detect lensing signatures. Using strong lensing effects, where background
galaxies are stretched into giant arcs and/or show up as multiple
images, it is possible to probe the cluster potential roughly out to the
Einstein radius. On larger scales, the shear and
depletion due to the lens may be used to constrain its
mass distribution, once these lensing effects are disentangled
from the intrinsic properties of the sources.

From simple arguments, it can be shown that shear detections have a
higher signal-to-noise ratio than magnification detections with the
number count method
(\cite{Mellier;1999,Bartelmann+Schneider;2000}). It must however be
emphasized, that from the observational point of view it is
easier to obtain deep galaxy number counts than detailed shear maps
that require good seeing and accurate determination of the PSF for
deconvolution. Note also, that ground based shear detections are
limited to galaxy
sizes above the seeing limit. With $0.5 \arcsec$ seeing in the $R$ band
this corresponds to \RAB $\leq 25$, giving a galaxy number density of
roughly $30$ arcmin$^{-2}$. Depletion analysis on the other hand
benefits from detections up to the magnitude limit, so that with deep
enough observations one can get much higher counts.
Another aspect in favor of the number count
method is that it is not subject to the ``sheet-mass degeneracy'' of the shear
analysis (e.g.~\cite{Schneider+King+Erben;2000}).
This does, however, not come for free, since
absolute surface density measurements using depletion require an accurate
determination of the unlensed background density
(\cite{Schneider+King+Erben;2000}). The intrinsic uncertainties due to
the combination of observational errors, including
contamination from foreground objects and false detections,
and clustering of the background population, certainly plague the
number count method.
In some wavelength bands these uncertainties make it difficult to
obtain significant constraints on model parameters
(\cite{Gray+Ellis+Refregier+;2000}).

Our data and analysis
show that detailed inferences from the present data are
to some extent
sensitive to poorly known a priori parameters ($\alpha$, $\bar{n}_0$,
$\eta$, \ldots)
making it difficult to set strong limits on the parameters of the lens
model. However, we do find compatible results for the two
bands when the slope of the differential number counts of the
background objects, $\alpha$, is lower than $0.25$
independent of the uncertainty in the unlensed background
densities. It is worth noting, that the $U$ band depletion curve
is both shallower
and narrower than the $R$ band depletion, suggesting a
steeper slope (\cite{Mayen+Soucail;2000}) in the $U$ band.
The two extreme values of the Einstein radius,
$\rE=17\arcsec$ and $\rE=25\arcsec$, found from the $U$ and $R$ band
data respectively for $\alpha \leq 0.25$,
may be used together with the
observed Einstein radius, $r_{\mrm{E,gal}}$, associated with the
single background galaxy
at $z=1.675$ (\cite{Broadhurst+Huang+Frye+;2000}) to estimate the
median, $\zM$, of the redshift distribution of
the background galaxies. In the case of a flat $\Omega=1$ cosmology,
we find that the ratio $\rE/r_{\mrm{E,gal}} = 17/30 \approx 0.57$
corresponds to
$\zM = 0.7$ while $\rE/r_{\mrm{E,gal}} = 25/30 \approx 0.83$
corresponds to $\zM = 1.1$. This latter value is in good agreement
with the models of \cite{Mayen+Soucail;2000}, although the
incompleteness of our data at faint magnitudes and the different
filters used make a direct comparison difficult. Note also, that
taking into account the uncertainty on the typical $U$ band Einstein
radius estimate of $\rE(U) = 20\arcsec \pm 3\arcsec$ and the $R$ band
estimate of $\rE(R) = 25\arcsec \pm 3\arcsec$ these two values do not
differ significantly.

The maximum likelihood analysis estimates of the Einstein radius using
a singular isothermal sphere model is consistent with the observed
giant arcs, provided that the slope $\alpha \leq 0.25$.
The estimated Einstein radius is independent of the dispersion of the
unlensed background density, parameterized by $\eta$, provided
that our estimate of the background density is biased low. We would
indeed expect that to be the case since the cluster is likely to
deplete the background density beyond the $150\arcsec$ assumed in our
estimation of the background density. We conclude that although it is
difficult to
draw strong conclusions about e.g.~the different galaxy
populations probed with the two bands without a more rigorous
treatment of the effects of a variable $\alpha$ and firmer
estimates of the unlensed background densities,
we do get results that are consistent with results obtained by
different methods by others.

One of the main aims of our work was to look for depletion of
background galaxies in the $U$ band. Our data clearly show this effect
towards the cluster CL0024+1654, giving independent evidence for a
flattening to a slope $\beta < 1$ ($\alpha < 0.4$) at the faint end
of the $U$ band luminosity function. 
While the reality of this flattening is still debatable
(e.g.~\cite{Pozzetti+Madau+;1998,Volonteri+Saracco+;2000}),
further depletion studies in the $U$ band will help to constrain
the slope of the number counts in this band.
It is clear that ultimately the shape of the number counts at faint
$U$ band magnitudes must be determined from deep, reliable observations
in the field. This would also settle the absolute unlensed
density of background sources, both of which is necessary for making
$U$ band depletion a reliable tool in cluster studies.

As the dominant cluster galaxies are relatively faint in the
$U$ band, we are able to measure the background number density close
to the cluster center accurately enough to find a significant turn-up
in the density as expected from simple lens models. This strengthens
the indication of a turn-up in the $R$ band, where the estimated
background density
is starting to increase before reaching the innermost parts of the
cluster where the foreground objects cover a too large fraction of the
available area to warrant a significant measurement.

With the present data, where the background density is of order
$30-50$ arcmin$^{-2}$, the inherent uncertainties of the number count
method still give rather loose bounds on lens model parameters, both
when fitting models to radially averaged depletion curves, and when more
elaborate maximum likelihood analysis is applied.
It goes without saying that with deeper number counts towards lenses,
and better estimates of the properties of the unlensed background
population, the reliability of the number count method will increase.
Some of the difficulties of the parametric modeling
inherent to these methods may however also be
alleviated by using more sophisticated, non-parametric methods.
We plan to study this in more detail for the data presented in this
paper.

\Myacknowledgements
This work was supported in part by the Danish Research Foundation
(through its establishment of the Theoretical Astrophysics Center),
The Icelandic Research Council, The Research Fund of the University of
Iceland, The Danish Natural Science Research Council (SNF),
and by the ``Access to Large-Scale Facilities''
program of the European commission of the EU.
The Nordic Optical Telescope is operated on the island of La Palma
jointly by Denmark, Finland, Iceland, Norway and Sweden, in the Spanish
Observatorio del Roque de los Muchachos of the Instituto de Astrofisica
de Canarias.
We are indebted to Ian Smail, David Hogg, and Lucia Pozzetti for
providing data on galaxy number counts, and to Oliver Czoske for
providing redshift data for CL0024+1654. We thank the referee for
useful comments and suggestions.
\Myacknowledgementsend

\references

\end{document}